# Econophysics of Business Cycles: Aggregate Economic Fluctuations, Mean Risks and Mean Square Risks

Victor Olkhov

TVEL, Kashirskoe sh. 49, Moscow, 115409, Russia

victor.olkhov@gmail.com

## Abstract

This paper presents hydrodynamic-like model of business cycles aggregate fluctuations of economic and financial variables. We model macroeconomics as ensemble of economic agents on economic space and agent's risk ratings play role of their coordinates. Sum of economic variables of agents with coordinate $\boldsymbol{x}$ define macroeconomic variables as functions of time and *coordinates* $\boldsymbol{x}$. We describe evolution and interactions between macro variables on economic space by hydrodynamic-like equations. Integral of macro variables over economic space defines aggregate economic or financial variables as functions of time $t$ only. Hydrodynamic-like equations define fluctuations of aggregate variables. Motion of agents from low risk to high risk area and back define the origin for repeated fluctuations of aggregate variables. Economic or financial variables on economic space may define statistical moments like mean risk, mean square risk and higher. Fluctuations of statistical moments describe phases of financial and economic cycles. As example we present a simple model relations between Assets and Revenue-on-Assets and derive hydrodynamic-like equations that describe evolution and interaction between these variables. Hydrodynamic-like equations permit derive systems of ordinary differential equations that describe fluctuations of aggregate Assets, Assets mean risks and Assets mean square risks. Our approach allows describe business cycle aggregate fluctuations induced by interactions between any number of economic or financial variables.

[1] This research did not receive any specific grant or financial support from TVEL or funding agencies in the public, commercial, or not-for-profit sectors.

# 1. Introduction

Fluctuations and waves are core properties of any complex system. Macroeconomic fluctuations are the most valuable processes that have impact on all characters of economic evolution and state. Modeling and forecasting business cycles aggregate fluctuations for decades remain focal point of economic studies [1-8]. Origin and drivers of aggregate fluctuations of economic and financial variables establish the key problems of business cycles. Let quote only three statements "Theories of business cycles should presumably help us to understand the salient characteristics of the observed pervasive and persistent non seasonal fluctuations of the economy". [3]. "Why aggregate variables undergo repeated fluctuations about trend, all of essentially the same character? Prior to Keynes' General Theory, the resolution of this question was regarded as one of the main outstanding challenges to economic research, and attempts to meet this challenge were called business cycle theory." [4]. "One of the most controversial questions in macroeconomics is what explains business-cycle fluctuations?" [8].

Our paper presents a general model of aggregate fluctuations of economic and financial variables and proposes further features that describe state and evolution of business cycles phases. Due to [6] "The real business cycle theory is a business cycle application of the Arrow-Debreu model, which is the standard general equilibrium theory of market economies". Our approach to business cycles is completely different from models based on assumptions of general equilibrium [9], decisions making [10] and behavioral economics [11]. We omit review of current state of business cycles and site [1-8] for great in-depth discussion and numerous references.

We regard business cycles as essential property of economic evolution and describe aggregate fluctuations of economic and financial variables as necessary feature of economic processes. Our model of aggregate fluctuations does not require existence any external shocks and disturbances of economic variables. Aggregate fluctuations reflect hidden evolution of economic and financial variables of economic agents on economic space [12-18]. Agent-based economic models are well known [19] but we suggest a different approach. Let assume that it is possible estimate risk rating for all agents of entire economics like huge banks and corporations and small firms and households. Let treat agent's risk ratings as their coordinates *alike to* coordinates of physical particles. Each economic agent has a lot of economic and financial variables as Assets and Debts, Credits and Loans, Production Function and Consumption and so on. Huge number of economic agents of entire economics can be treated *alike to* "economic gas". Certain parallels to kinetic theory of gases allows establish transition from description of economic variables of separate agents with coordinate $\boldsymbol{x}$ to description of economic variables defined as cumulative variables of all agents with coordinate $\boldsymbol{x}$. Such transition has parallels to transfer from description of kinetic multi particle system to hydrodynamic approximation that neglect particle granulation and describe system as a continuous media. Similar considerations allow develop transition from description of macroeconomics as system of numerous separate economic agents with economic and financial variables to macroeconomic model that neglect agent's granularity and describe economic and financial variables as functions of time and coordinates on economic space alike to continuous media or hydrodynamic approximation in physics. Vital distinctions between economics and physical systems prohibit any "deep analogies". Nevertheless certain parallels between multi-agent systems on economic space and multi-particle systems on physical space allow derive hydrodynamic-like equations that describe evolution of cumulative economic and financial variables. Such approach to macroeconomic modeling makes visible and describes a wide range of internal economic and financial waves that are induced by perturbations of variables on economic space [12-18].

In this paper we show that our model can describe business cycle aggregate fluctuations of economic and financial variables. Hydrodynamic-like equations describe dynamics of economic and financial variables as functions of time and coordinates on economic space. Integrals of economic variables over space coordinates define aggregate economic and financial variables as functions of time only. For example integral by Assets $A(t,\boldsymbol{x})$ over economic space defines aggregate Assets $A(t)$ as function of time only:

$$A(t) = \int d\boldsymbol{x}\; A(t,\boldsymbol{x})$$

Usage of hydrodynamic-like equations allow derive ordinary differential equations on aggregate variables and model their growth and fluctuations in time. Non-linear character of hydrodynamic-like equations induces chains of factors that could govern growth and fluctuations of aggregate economic and financial variables. Our approach reveals that observed aggregate fluctuations may depend not on other economic variables but on various hidden economic factors that define evolution of economic system on economic space. For example we show that aggregate fluctuations of macroeconomic Assets $A(t)$ may depend on factor that have meaning *alike to* energy $EA(t)$ of Assets flow in hydrodynamics - $EA(t)$ is proportional to Assets density $A(t,\boldsymbol{x})$ and square of velocity of Assets flow on economic space.

Usage of economic space with coordinates that have meaning of agent's risk ratings allows describe business cycles not only as aggregate fluctuations but by additional new parameters. Most economic and financial variables are positive defined functions on economic space. Each of positive distribution can be used as probability distribution and can define mean risks or mean coordinates of particular economic variable. For example distribution of Assets $A(t,\boldsymbol{x})$ on economic space can define Assets mean risk $\boldsymbol{X}(t)$ as

$$\boldsymbol{X}(t)A(t) = \int d\boldsymbol{x}\ \boldsymbol{x}\,A(t,\boldsymbol{x})$$

As we show below, hydrodynamic-like equations on Assets distribution $A(t,\boldsymbol{x})$ allow derive equations that describe dynamics and fluctuations of $\boldsymbol{X}(t)A(t)$. In this paper for simple model we show that time fluctuations of aggregate Assets $A(t)$ and action of additional factors define fluctuations of Assets mean risks $\boldsymbol{X}(t)$ with frequencies that can be different from frequencies of aggregate Assets $A(t)$ fluctuations. We propose that economic phases or business cycles should be characterized my many parameters and aggregate fluctuations are only the simplest ones. For example Assets mean square risk $\overline{X^2}(t)$

$$\overline{X^2}(t)A(t) = \int d\boldsymbol{x}\ x^{\mathbf{2}}\,A(t,\boldsymbol{x})$$

defines second statistical moment of Assets distribution on economic space and determines its rate of risk uncertainty due to Assets dispersion $\sigma^2(x)$

$$\sigma^2(x) = \overline{X^2}(t) - X^2(t)$$

Growth of assets dispersion indicates rise of macroeconomic Assets risk uncertainties and decrees of dispersion reflect clustering of macroeconomic Assets near mean Assets mean risk ***X****(t)*. Description of mean risks, mean square risks and further statistical moments for different economic and financial variables can describe economic phases and business cycles in greater detail. Different economic and financial variables induce different values of mean risks, mean square risks and etc. Description of aggregate fluctuations of economic and financial variables requires taking into account fluctuations of other aggregate variables and factors that are not observed up now. For example, modeling of Assets $A(t)$ aggregate fluctuations may require factor that have *the appearance* of Assets energy $EA(t)$

$$EA(t) = \int dx\,EA(t,\boldsymbol{x}) = \int dx\ A(t,\boldsymbol{x})\,v^2(t,\boldsymbol{x})$$

Energy $EA(t,\boldsymbol{x})$ of the Assets $A(t,\boldsymbol{x})$ is proportional to Assets density $A(t,\boldsymbol{x})$ and square of Assets velocity $v^2(t,\boldsymbol{x})$ on economic space *alike to* energy of hydrodynamic flow but no parallels to physical energy exist. Such unexpected factors can induce major impact on business cycle aggregate fluctuations and their possible influence should be studied further.

Economic origin of aggregate fluctuations described by our model is very "simple". There is no need in any perturbations or external shock to derive equations and describe aggregate fluctuations. Economic evolution induces motions of economic agents from low risk affairs to high risk business and back and that motion of agent's risks cause corresponding flows and tides of economic and financial variables on economic space. Such tides of economic and financial variables from low to high risks areas and back govern growth and aggregate fluctuations of all economic and financial variables as GDP, Investment, Assets and so on. Enormous number of economic and financial variables and complexity of their mutual interactions makes that "simple" problem extremely difficult.

In this paper we present simplest model that describe mutual interactions between two variables only - between Assets $A(t,\boldsymbol{x})$ and Revenue-on-Assets $B(t,\boldsymbol{x})$ on economic space. We derive equations that describe growth and fluctuations of aggregate Assets $A(t)$ and aggregate Revenue-on-Assets $B(t)$. We derive equations on Assets mean risks $\boldsymbol{X}(t)$ and mean square risks $\overline{X^2}(t)$ and describe their fluctuations also.

The rest of the paper is organized as follows. In Section 2 we argue the model setup and remind major issues for economic space modeling [12-14]. We argue multi agent systems and present reasons for usage of hydrodynamic-like equations. In Section 3 we discuss business cycle aggregate fluctuations in terms of economic space model. We explain dependence of aggregate fluctuations on description of economic distributions by hydrodynamic-like equations on economic space. As example we present a simple model and hydrodynamic-like equations that describe mutual interactions between Assets $A(t,\boldsymbol{x})$ and Revenue-on-Assets $B(t,\boldsymbol{x})$ economic variables. We show that solutions of aggregate Assets $A(t)$ and Revenue-on-Assets $B(t)$ follow exponential growth and fluctuations in time. We derive equations that describe Assets mean risks $\boldsymbol{X}(t)$ and Revenue-on-Assets mean risks $\boldsymbol{Y}(t)$ and show that they depend on factors different from those determine aggregate Assets $A(t)$ and their solutions follow time fluctuations also but with different frequency. The same considerations can be applied to any aggregate economic and financial variables as GDP, Investment, Demand and etc. Conclusions are in Section 4. In Appendix A we derive system of ordinary differential equations that describe time fluctuations of aggregate Assets $A(t)$ and aggregate Revenue-on-Assets $B(t)$. Appendix B gives derivation of ordinary differential equations that describe time fluctuations of Assets mean risks $\boldsymbol{X}(t)$ and Revenue-on-Assets mean risks $\boldsymbol{Y(t)}$. Appendix C presents derivation of equations on Assets mean square risks $\overline{X^2}(t)$.

## 2. Model Setup

Extreme complexity of macroeconomic and financial processes indicates that their description via time-series analysis of economic variables might be not sufficient for development of adequate models and forecasts. We propose that economic modeling should be based on description of economic and financial variables treated as functions of time and coordinates on certain space that reflect essential economic properties. Economics is so different from natural processes that no physical or geographical space can play such a role and our approach has nothing common with spacial economics [20]. We propose to use well-known economic issues but transform them in such a way to uncover and outline their properties of "economic space". For decades macroeconomics and finance measure risks associated with huge banks and corporations by their risk ratings issued by international rating agencies [21-23]. Let make several suppositions. Let assume that current risks assessment methodology can be extended and generalized in order to estimate risk ratings as for major banks and corporations as for all agents of entire economics. Let propose that it is possible to estimate ratings for all risks that may affect macroeconomic and financial evolution. If so let use agent's risk ratings as their coordinates. Let call such a space that imbed agent's risk ratings as coordinates – economic space. Usage of such well known economic issues as risk ratings allow describe relations between economic and financial variables by partial differential equations and that uncovers a hidden complexity of economic evolution processes. This approach allows model concealed origin of business cycle aggregate fluctuations of economic and financial variables and describe other features that characterize states of economic evolution phases. Below we introduce economic space modeling relations according to [12-18].

### *2.1. Economic Space*

Let regard extensive (additive) macroeconomic variables as Assets, Debts, Production Function, Consumption and Investment, Credits and Loans and etc. All macroeconomic and financial variables are determined as aggregates of corresponding variables of economic agents. Assets of entire economics are defined by composition of agent's Assets. Aggregative Production Functions of economic agents define macroeconomic Production Function. Economic growth and fluctuations of macroeconomic and financial variables are determined by economic growth and fluctuations of corresponding variables of economic agents. To describe evolution of economic and financial variables one should model corresponding evolution of economic agents. We propose that description of macroeconomic and financial time-series is not sufficient to model drivers and hidden interactions that govern economic evolution. To develop adequate models of evolution of economic agents and their influence on dynamics of macroeconomic and financial variables we introduce economic space notion [12-18]. Our approach is completely different from general equilibrium [9], economic decision-making [10], behavioral economics [11], agent-based economics [19] and spatial economics [20].

We propose that business cycles observed as time fluctuations of macroeconomic and financial variables are induced by a concealed dynamics and wave propagation of economic variables. Description of waves of economic and financial variables requires a certain space where economic waves can propagate. Introduction of essential economic space as ground for modeling agents evolution allows describe dynamics of agent's economic and financial variables *alike to* description of multi particle systems. We outline vital differences between economic and physical systems but demonstrate that usage of similar concepts allows develop useful parallels between description of continuous media in physics and modeling business cycles, growth and fluctuations of macroeconomic and financial variables.

Main issue of economic space notion is simple. To describe dynamics of numerous economic agents we propose use agent's risk ratings as their coordinates on economic space. We don't study ratings of specific risks like credit, liquidity, market risks and propose regard ratings of any risks that can distress agents and hence macroeconomics and finance as coordinates of agents on economic space. Such simple proposition hides many problems. Up now risk ratings are provided by international rating agencies [21-23] not for all economic agents but for major banks and corporations only. Thus to establish our model let assume that generalization of risk methodologies and development of available econometric data can make possible assessment of risk ratings for all economic agents of macroeconomic system - for huge corporations and for small firms and even for households, and for any risks that may affect evolution of macroeconomic and finance. Below we present brief reasons for economic space definition due to [12-14].

International rating agencies [21-23] estimate risk ratings of huge corporations and banks and these ratings are widespread in current economics and finance. Risk ratings take values of risk grades and noted as *AAA, BB, C* and so on. Let treat such risk grades like *AAA, BB, C* as points $x_1, x_2,.. x_m$ of discreet space. Let propose, that risk assessments methodologies can be extended to estimate risk ratings for all agents of entire economics. That will distribute all agents over points of finite discreet space determined by set of risk grades. Many risks impact macroeconomics. Let regard grades of single risk as points of one-dimensional space and simultaneous assessments of *n* different risks as measurements of coordinates of agents on *n*-dimensional economic space. Let propose, that risk assessments methodologies can be generalized in such a way that risk grades can take continuous values and define space *R*. Thus risk grades of *n* different risks establish $R^n$.

Modeling on economic space uncovers extreme complexity of macroeconomics and finance and can't take into account all possible risks. Description of agents of entire macroeconomics on economic space $R^n$ requires choice of *n* major risks that cause major effects on economic and financial processes. To determine economic space one should estimate current risks and select two, three, four major risks as main factors affecting economic system. That establishes economic space with two or three dimensions. To select most valuable risks one should compare impact of different risks on economic and financial processes and chose few most valuable. Selection of *n* major risks defines initial representation of economic space $R^n$ and that is a separate and very tough problem.

It is well known that risks can suddenly arise and then vanish. To describe evolution of economic system and its agents in a time term *T* one should forecast *m* main risks that can play major role in a particular time term. Such forecast define target state of economic space $R^m$ in a time term *T*. To describe evolution of economics taking into account variations of major risks during time term *T* one should define transition from initial economic space $R^n$ determined by *n* main risks to target economic space $R^m$ determined by *m* main risks. Such transitions describe how initial set of *n* risks can decline its action on economic agents and how new *m* risks grow up. Transition from initial set of *n* main risk to target set of *m* risks describes evolution from initial economic space $R^n$ of to the target one $R^m$. Such unpredictable changes of major risks and corresponding changes of economic space representation are origin of irremovable randomness of economic and financial evolution.
Selection of main risks simplifies description and allows neglect "small risks". Selections of major risks give opportunity to validate initial and target sets of risks and to prove or disprove initial model assumptions. It makes possible to compare predictions of economic model with real data and outlines disagreements between predictions and observations.

Below we develop models on economic space $R^n$ in the simple assumption that macroeconomics and economic agents are under permanent action *n* risks. We don't study transitions from one set of major risks to another and describe simple models on constant economic space $R^n$. Agent's risk ratings play role of their coordinates on economic space $R^n$. Introduction of economic space allows describe evolution of economic agents *alike to* description of multi-particle systems. We repeat that distinctions between economics and physics are absolutely vital but certain parallels between them allow develop economic models *similar to* description of multi-particle systems and hydrodynamics.

## *2.2 Multi-agent system*

Introduction of economic space allows substitute widespread partition of agents by economic sectors and industries with partition of agents by their coordinates on economic space [12-15]. Decomposition of economics by sectors allows define Assets or Profits of Bank sector as cumulative Assets or Profits of all agents of this particular sector. Let replace decomposition of economics by sectors and let allocate agents by their risk ratings $\boldsymbol{x}$ as coordinates $\boldsymbol{x}$ on economic space. Such allocation allows define macroeconomic variables as functions of $\boldsymbol{x}$ on economic space. For example, cumulative Assets of all agents with coordinate $\boldsymbol{x}$ define macroeconomic Assets as function of coordinate $\boldsymbol{x}$. Such approach allows describe economic and financial processes on economic space *alike to* description of multi-particle system in physics in the continuous media or hydrodynamic approximation. Indeed, agents risk ratings $\boldsymbol{x}$ or agents coordinates $\boldsymbol{x}$ change under the action of economic and financial processes. Agents move on economic space *alike to* economic particles or "*economic gas*". Motion of agents on economic space causes change of agent's economic and financial variables. Let describe agents and their variables by probability distributions. Averaging of agent's economic and financial variables by probability distributions allow describe economics alike to continuous media or hydrodynamic-like approximation. In such approximation we neglect granularity of variables like Assets or Capital that belong to separate agents at point $\boldsymbol{x}$ and describe macroeconomic Assets or Capital as function of $\boldsymbol{x}$ on economic space alike to "*Assets fluid*" or "*Capital fluid*" in hydrodynamics. In some sense such transition has parallels to partition of Assets by sectors or industries. The "small" difference: in common approach agents and their variables belong to permanent industry or sector. In our model agent's risk ratings define *linear space* and *agents can move* on economic space due to change of their risk ratings. These small distinctions allow model economics as a "*continuous economic media*".

Below for convenience we present definition of macroeconomic and financial variables according to [12-18]. For brevity let further call economic agents as economic particles or *e-particles* and economic space as *e-space*. Let introduce macro variables at point $\boldsymbol{x}$ as sum of variables of e-particles with coordinates $\boldsymbol{x}$ on e-space.

Each e-particle has many economic and financial variables like Assets and Debts, Investment and Savings, Credits and Loans, Production Function and Consumption and etc. Let call e-particles as "*independent*" if sum of extensive (additive) variables of any group of e-particles equals same variable of entire group. For example: sum of Assets of *n* e-particles equals Assets of entire group. Let assume that all e-particles are "*independent*" and sum of extensive variables of any group of agents equals same variable of entire group. For example, aggregation of Assets of e-particles with coordinates $\boldsymbol{x}$ on e-space define Assets as function of time *t* and $\boldsymbol{x}$. Integral of Assets $A(t,\boldsymbol{x})$ by $d\boldsymbol{x}$ over e-space equals aggregate Assets $A(t)$ of entire macroeconomics. We mention Assets as example of macroeconomic variable only and our considerations valid for any extensive economic or financial variable. Below we show that description of dynamics of economic and financial variables as functions of time *t* and coordinate $\boldsymbol{x}$ on economic space allows model business cycles, growth and fluctuations of macro variables of entire economics as functions of time *t*.

Coordinates of e-particles represent their risk ratings and hence they are under random motion on e-space. Thus sum of Assets of e-particles near point $\boldsymbol{x}$ is random also. To obtain regular values of macro variables like Assets at point $\boldsymbol{x}$ let average Assets at point $\boldsymbol{x}$ by probability distribution *f*. Let state that distribution *f* define probability to observe $N(\boldsymbol{x})$ e-particles with value of Assets equal $a_1,...a_{N(x)}$. That determine density of Assets at point $\boldsymbol{x}$ on e-space (Eq.(2.1) below). Macro Assets as function of time *t* and coordinate $\boldsymbol{x}$ behave *alike to* Assets fluid - *similar to* fluids in hydrodynamics. To describe motion of Assets fluid [12] let define velocity of such a fluid. Let mention that velocities of e-particles are not additive variables and their sum doesn't define velocity of Assets motion. To define velocities of Assets fluid correctly one should define "*Asset's impulses*" $\boldsymbol{p}_j$ at point $\boldsymbol{x}$ as product of Assets $a_j$ of particular *j*-e-particle and its velocity $\boldsymbol{v}_j$ (Eq. (2.2) below). Such "Asset's impulses" $\boldsymbol{p}_j = a_j \boldsymbol{v}_j$ - are additive variables and sum of "Asset's impulses" can be averaged by similar probability distribution *f*. Densities of Assets and densities of Assets impulses permit define velocities of Assets fluid (Eq.(2.3) below). Different economic and financial fluids can flow with different velocities. For example flow of Capital on e-space can have velocity higher then flow of Assets, nevertheless they are determined by motion of same e-particles. Macroeconomics can be modeled as interaction between numerous economic fluids and that makes description extremely difficult. Let present these issues in a more formal way.

Let assume that each e-particle on e-space $R^n$ at moment $t$ is described by $l$ extensive variables $(u_1,\ldots u_l)$. Extensive variables are additive and admit averaging by probability distributions. Intensive variables, like Prices or Interest Rates, cannot be averaged directly. Enormous number of extensive variables like Capital and Credits, Investment and Assets, Profits and Savings, etc., describe each e-particle and make economic modelling very complex. As usual, macro variables are defined as aggregates of corresponding values of all e-particles of entire economics. For example, macro Investment equal aggregate Investment and Assets can be calculated as cumulative Assets of all e-particles. Let define macro variables as functions of time $t$ and coordinates $\boldsymbol{x}$ on e-space.

Let assume that there are $N(\boldsymbol{x})$ e-particles at point $\boldsymbol{x}$. Let state that velocities of e-particles at point $\boldsymbol{x}$ equal $\boldsymbol{v}=(\boldsymbol{v}_1,\ldots \boldsymbol{v}_{N(x)})$. Each e-particle has $l$ extensive variables $(u_1,\ldots u_l)$. Let assume that values of variables equal $u=(u_{1i},\ldots u_{li})$, $i=1,..N(\boldsymbol{x})$. Each extensive variable $u_j$ at point $\boldsymbol{x}$ defines macro variable $U_j$ as sum of variables $u_{ji}$ of $N(\boldsymbol{x})$ e-particles at point $\boldsymbol{x}$

$$U_j = \sum_i u_{ji}\,; \quad j = 1,..l; \quad i = 1,\ldots N(\boldsymbol{x})$$

To describe motion of variable $U_j$ let establish additive variable alike to impulse in physics. For e-particle $i$ let define impulses $p_{ji}$ (1.1) as product of extensive variable $u_j$ that takes value $u_{ji}$ and its velocity $\boldsymbol{v}_i$:

$$p_{ji} = u_{ji}\boldsymbol{v}_i \tag{1.1}$$

For example if Assets $a$ of e-particle $i$ take value $a_i$ and velocity of e-particle $i$ equals $\boldsymbol{v}_i$ then impulse $pa_i$ of Assets of e-particle $i$ equals $pa_i = a_i \cdot \boldsymbol{v}_i$. Thus if e-particle has $l$ extensive variables $(u_1,\ldots u_l)$ and velocity $\boldsymbol{v}$ then it has $l$ impulses $(p_1,p_2,..p_l)=(u_1 \cdot \boldsymbol{v},\ldots u_l \cdot \boldsymbol{v})$. Let define impulse $\boldsymbol{P}_j$ (1.2) of variable $U_j$ as

$$\boldsymbol{P}_j = \sum_i u_{ji} \cdot \boldsymbol{v}_i\,; \quad j = 1,..l; \quad i = 1,\ldots N(\boldsymbol{x}) \tag{1.2}$$

Let introduce economic distribution function $f=f(t,x;U_1,..U_l, \boldsymbol{P}_1,..\boldsymbol{P}_l)$ that determine probability to observe variables $U_j$ and impulses $\boldsymbol{P}_j$ at point $\boldsymbol{x}$ at time $t$. $U_j$ and $\boldsymbol{P}_j$ are determined by corresponding values of e-particles that have coordinates $\boldsymbol{x}$ at time $t$. They take random values at point $\boldsymbol{x}$ due to random motion of e-particles on e-space. Averaging of $U_j$ and $\boldsymbol{P}_j$ within economic distribution function $f$ allows establish transition from approximation that takes into account variables of separate e-particles to continuous "*economic media*" or hydrodynamic-like approximation that neglect e-particles granularity and describe averaged macro variables as functions of time and coordinates on e-space. Let define economic or financial density functions

$$U_j(t,\boldsymbol{x}) = \int U_j\, f(t,\boldsymbol{x},U_1,\ldots U_l,\boldsymbol{P}_1,..\boldsymbol{P}_l)\, dU_1..dU_l d\boldsymbol{P}_1..d\boldsymbol{P}_l \tag{2.1}$$

and impulse density functions $\boldsymbol{P}_j(t,x)$

$$\boldsymbol{P}_j(t,\boldsymbol{x}) = \int \boldsymbol{P}_j\, f(t,\boldsymbol{x},U_1,\dots U_l,P_1,..P_l)\, dU_1..dU_l d\boldsymbol{P}_1..d\boldsymbol{P}_l \quad (2.2)$$

That allows define e-space velocities $\boldsymbol{v}_j(t,\boldsymbol{x})$ (2.3) of densities $U_j(t,\boldsymbol{x})$ as

$$U_j(t,\boldsymbol{x})\boldsymbol{v}_j(t,\boldsymbol{x}) = \boldsymbol{P}_j(t,\boldsymbol{x}) \quad (2.3)$$

Densities $U_j(t,\boldsymbol{x})$ (2.1) and impulses $\boldsymbol{P}_j(t,\boldsymbol{x})$ (2.2) are determined as mean values of aggregates of corresponding variables of separate e-particles with coordinates $\boldsymbol{x}$. Functions $U_j(t,\boldsymbol{x})$ can describe macro densities of Investment and Loans, Assets and Debts and so on. For example, Assets density $A(t,\boldsymbol{x})$, impulse $\boldsymbol{P}(t,\boldsymbol{x})$ and velocity $\boldsymbol{v}(t,\boldsymbol{x})$ can be defined as

$$A(t,\boldsymbol{x}) = \int a\, f(t,\boldsymbol{x},a,\boldsymbol{P})\, dad\boldsymbol{P} \quad (2.4)$$

$$\boldsymbol{P}(t,\boldsymbol{x}) = \int \boldsymbol{P}\, f(t,\boldsymbol{x},a,\boldsymbol{P})\, dad\boldsymbol{P} \quad (2.5)$$

$$A(t,\boldsymbol{x})\boldsymbol{v}(t,\boldsymbol{x}) = \boldsymbol{P}(t,\boldsymbol{x}) \quad (2.6)$$

Here $a$ and $\boldsymbol{P}$ denote sum of Assets and impulses of all e-particles with coordinates $\boldsymbol{x}$. To describe evolution of macro densities like Investment and Loans, Assets and Debts and etc., let derive hydrodynamic-like equations and use Assets density $A(t,\boldsymbol{x})$, impulse $\boldsymbol{P}(t,\boldsymbol{x})$ and velocity $\boldsymbol{v}(t,\boldsymbol{x})$ as example.

## 2.3. Hydrodynamic-like equations

In this section we present hydrodynamic-like equations for economic and financial densities like Capital and Assets, Investment and Credits and etc. [12-18]. Let follow [24] and for any extensive economic or financial density like Assets $A(t,\boldsymbol{x})$ (2.4), its impulse $\boldsymbol{P}(t,\boldsymbol{x})$ (2.5) and its velocity $\boldsymbol{v}(t,\boldsymbol{x})$ (2.6) on e-space hydrodynamic-like equations take form:

$$\frac{\partial A}{\partial t} + \nabla \cdot (\boldsymbol{v}A) = Q_1 \quad (3.1)$$

Left hand side describes two factors that can change value of Assets $A(t,\boldsymbol{x})$ in a unit volume on e-space. First factor $\frac{\partial A}{\partial t}$ describes change of $A(t,\boldsymbol{x})$ in time. Second factor $\nabla \cdot (\boldsymbol{v}A)$ describes change of $A(t,\boldsymbol{x})$ due to flux $\boldsymbol{v}A$ through surface of unit volume according to the Gauss-Ostrogradsky theorem: the integral of the divergence $\nabla \cdot (\boldsymbol{v}A)$ over volume $V$ equals the surface integral of flux $\boldsymbol{v}A$ over the boundary of volume $V$. These two factors describe possible change of Assets density $A(t,\boldsymbol{x})$ in a unit volume. Right hand side factor $Q_1$ describes any action of other factors or densities that can change left side. Assets impulse $\boldsymbol{P}(t,\boldsymbol{x})$ follows similar equations:

$$\frac{\partial \boldsymbol{P}}{\partial t} + \nabla \cdot (\boldsymbol{v}\boldsymbol{P}) = \boldsymbol{Q}_2 \quad (3.2)$$

Right hand side factor $\boldsymbol{Q_2}$ describes any action of other densities that can change left side. Parallels to equations of hydrodynamics are formal and extrinsic. We repeat that intrinsic, essential relations of economics and finance have nothing common with physical laws. Hydrodynamic-like equations (3.1; 3.2) describe very simple relations: left side describes to possible factors that can change amount of any extensive (additive) density in a unit volume: due to change in time and due to flux through unit surface. Right side describes any external factors that can change left side. Thus economic meaning of hydrodynamic-like equations (3.1; 3.2) is defined by right hand side factors $Q_1$ and $\boldsymbol{Q_2}$. Due to parallels with hydrodynamics let call equations on densities (3.1) as Continuity Equations and equations on impulses (3.2) as Equations of Motion. It seems that equations (3.1; 3.2) are simple. But one should remember that enormous number of macroeconomic and financial variables and diversity of possible interactions between them makes entire problem much more complex then any problem of physical hydrodynamics.

To define factors $Q_1$ and $\boldsymbol{Q_2}$ let outline the following. E-particles (economic agents) don't collide in e-space *alike to* physical particles and agent's variables don't obey any conservation laws *alike to* conservation laws of mass, impulse, energy etc. Factors $Q_1$ and $\boldsymbol{Q_2}$ don't take into account any elements *alike to* viscosity, pressure and etc. Let state that factors $Q_1$ and $\boldsymbol{Q_2}$ in the right hand of hydrodynamic-like equations (3.1; 3.2) on any density $A(t,\boldsymbol{x})$ and impulses $\boldsymbol{P}(t,\boldsymbol{x})$ depend on densities $U_j(t,\boldsymbol{x})$ and impulses $\boldsymbol{P}_j(t,\boldsymbol{x})$ different from $A(t,\boldsymbol{x})$ and $\boldsymbol{P}(t,\boldsymbol{x})$. Let call such variables $U_j(t,\boldsymbol{x})$, $\boldsymbol{P}_j(t,\boldsymbol{x})$ as *conjugate* to variables $A(t,\boldsymbol{x})$ and $\boldsymbol{P}(t,\boldsymbol{x})$ if variables $U_j(t,\boldsymbol{x})$, $\boldsymbol{P}_j(t,\boldsymbol{x})$ determine $Q_1$ and $\boldsymbol{Q_2}$ factors in right hand side of hydrodynamic-like equations on $A(t,\boldsymbol{x})$ and $\boldsymbol{P}(t,\boldsymbol{x})$. For example, Investment may have *conjugate* variables like Cost of Capital or Return on Investment and their velocities. Demand may be *conjugate* to Supply and vice versa. Let state, that *conjugate* variables define right hand side of Continuity Equation and Motion Equations (3.1, 3.2). As we mentioned above, economic densities like Assets $A(t,\boldsymbol{x})$ are determined by corresponding variables of e-particles (economic agents) with coordinates $\boldsymbol{x}$. Any agent's variables change due to economic and financial transactions between agents. Thus $Q_1$ and $\boldsymbol{Q_2}$ factors that model action of *conjugate* variables should reflect relations defined by transactions between e-particles.

In this paper for simplicity we describe *a local approximation* of economic and financial transactions between e-particles that takes into account transactions between e-particles with same coordinates only. Such simplification [12-15, 17] allows describe factors $Q_1$ and $\boldsymbol{Q_2}$ by simple linear differential operators on *conjugate* densities. We use this assumption in the next Section. More complex models with *non-local approximations* that take into account "*action-at-a-distance*" transactions between e-particles (economic agents) on economic space are presented in [16, 18].

## 3. Business Cycles

Let try respond to question: "Why aggregate variables undergo repeated fluctuations about trend, all of essentially the same character?" [4].

We propose that [1-8] present sufficiently reasonable general considerations for business cycle motivations. We simply present a model that describes business cycle aggregate fluctuations of economic and financial variables about growth trend. We develop hydrodynamic-like models that describe evolution of economic and financial densities on economic space determined by dynamics of cumulative risk ratings. We don't argue problem: *why it happens*? We describe - *what happens*. Distribution of economic agents by their risk ratings as coordinates on economic space allows describe "repeated fluctuations about trend". Definition of economic and financial densities (2.1-2.6) as functions of time $t$ and coordinate $\boldsymbol{x}$ allows define variables of the entire macroeconomics as integrals by $d\boldsymbol{x}$ over e-space. For example integral (4.1)

$$A(t) = \int d\boldsymbol{x}\ A(t,\boldsymbol{x}) \tag{4.1}$$

of Assets density $A(t,\boldsymbol{x})$ over e-space defines aggregate Assets $A(t)$ of entire economics. Same relations define all aggregate economic and financial variables of entire economics as function of time $t$ only. Aggregate GDP, Investment, Credits, Taxes etc., can be presented similar to (4.1). Thus evolution of macroeconomic variables in time is determined by hidden dynamics of corresponding economic and financial densities on e-space. For example, evolution of macroeconomic aggregate Assets $A(t)$ in time is determined by dynamics of Assets density $A(t,\boldsymbol{x})$ on e-space. Economic growth and time fluctuations of all macroeconomic and financial variables are determined by evolution of densities on e-space. Economic densities at point $(t,\boldsymbol{x})$ are determined by variables of agents at the same point. Thus business cycles treated as "fluctuations about trend" of macroeconomic and financial variables are determined by dynamics of agent's variables under their motion on economic space.

Usage of e-space enhances methods for economic modeling and allows indicate new factors that describe state and evolution of entire economics. Indeed, we use agent's risk ratings as their coordinates on e-space. For each economic or financial density like Assets density $A(t,\boldsymbol{x})$ let define mean risk $\boldsymbol{X}(t)$ of particular density as

$$\boldsymbol{X}(t)A(t) = \int d\boldsymbol{x}\ \boldsymbol{x}\,A(t,\boldsymbol{x}) \tag{4.2}$$

Mean risk $\boldsymbol{X}(t)$ (4.2) reflects distribution of Assets by risk ratings – by coordinates of e-space. If agents and their Assets move to more risky domain then Assets density $A(t,\boldsymbol{x})$ shifts to more risky values and same happens with $\boldsymbol{X}(t)$. When agents run from risks and move to safe area then their Assets and Assets density $A(t,\boldsymbol{x})$ as well shift to lower risks and that induce corresponding motion of mean Assets risk $\boldsymbol{X}(t)$. So, Assets mean risk $\boldsymbol{X}(t)$ follows time fluctuations near certain stationary state on e-space and such fluctuations describe phases of economic evolution. We propose that fluctuations of mean risks can describe phases of business cycles. Mean risk $\boldsymbol{X}(t)$ is a first statistical moment of probability distribution $p(t,x)$ defined by economic or financial density. For example, Assets density $A(t,\boldsymbol{x})$ can define Assets probability distribution $p_A(t,x)$:

$$p_A(t,\boldsymbol{x}) = A^{-1}(t)A(t,\boldsymbol{x}) \ \ ; \int d\boldsymbol{x} \ p_A(t,\boldsymbol{x}) = 1 \tag{4.3}$$

and (4.3) define mean risk of Assets density:

$$\boldsymbol{X}(t) = \int d\boldsymbol{x} \ \boldsymbol{x} \, p_A(t,\boldsymbol{x}) \tag{4.4}$$

Width of distribution $p_A(t,x)$ reflects width of Assets density on e-space and second moment of distribution $p_A(t,x)$

$$\overline{X^2}(t) = \int d\boldsymbol{x} \ x^{\boldsymbol{2}} \, p_A(t,\boldsymbol{x}) \ ; \ \sigma^2(x) = \overline{X^2}(t) - X^{\boldsymbol{2}}(t) \tag{4.5}$$

Mean square $\overline{X^2}(t)$ and dispersion $\sigma^2(x)$ also should follow time fluctuations and can be useful for description business cycle phases. Growth of dispersion $\sigma^2(x)$ describes rise of economic risk uncertainty and $\sigma^2(x)$ reduction reflects clustering of economic agents and economic densities near mean risk value. Let show how hydrodynamic-like equations (3.1; 3.2) allow describe evolution and aggregate fluctuations of macroeconomic variables and mean risks. To do that let define factors $Q_1$ and $\boldsymbol{Q}_2$ for simple model of two *self-conjugate* economic densities.

## ***3.1. Model equations***

As example let describe relations between two extensive (additive) variables as Assets density $A(t,\boldsymbol{x})$ and Revenue-on-Assets $B(t,\boldsymbol{x})$. Returns on Assets are more widespread variable but it is intensive (not additive) variable. Revenue-on-Assets is extensive and additive variable and relations between these variables are well known. Let simplify the problem and propose that Revenue-on-Assets $B(t,\boldsymbol{x})$ and its impulse $\boldsymbol{PB}(t,\boldsymbol{x})$ are only *conjugate* variable that defines $Q_1$ and $\boldsymbol{Q}_2$ factors for equations (3.1; 3.2) on Assets $A(t,\boldsymbol{x})$ and its impulse $\boldsymbol{PA}(t,\boldsymbol{x})$ and vise versa: Assets $A(t,\boldsymbol{x})$ and its impulse $\boldsymbol{PA}(t,\boldsymbol{x})$ are only *conjugate* variable that defines $Q_1$ and $\boldsymbol{Q}_2$ factors for equations (3.1; 3.2) on Revenue-on-Assets $B(t,\boldsymbol{x})$ and its impulse $\boldsymbol{PB}(t,\boldsymbol{x})$. Let take simplest relations between two *self-conjugate* economic variables and that allow describe their mutual interaction in self-consistent manner. To simplify the model let propose that Revenue-on-Assets $B(t,\boldsymbol{x})$ and its impulse in moment $t$ depend on Assets and its impulse at same moment $t$ and vice versa. Let assume that factor $Q_{1A}$ for Continuity Equation (3.1) on Assets density $A(t,\boldsymbol{x})$ is proportional to scalar product between Revenue-on-Assets impulse $\boldsymbol{PB}(t,\boldsymbol{x})$ and vector $\boldsymbol{x}$, so:

$$Q_{1A} = a\,\boldsymbol{x} \cdot \boldsymbol{PB}(t,\boldsymbol{x}) \tag{5.1}$$

and same relations define factor $Q_{1B}$ for Continuity Equation (3.1) on Revenue-on-Assets density $B(t,\boldsymbol{x})$

$$Q_{1B} = b\,\boldsymbol{x} \cdot \boldsymbol{PA}(t,\boldsymbol{x}) \tag{5.2}$$

Here $a$ and $b$ – const. Economic meaning of (5.1; 5.2) is follows. Assets density in unit volume a point $(t,\boldsymbol{x})$ grow up if scalar product of Revenue-on-Assets flow $\boldsymbol{PB}(t,\boldsymbol{x}) = \boldsymbol{u}(t,\boldsymbol{x})B(t,\boldsymbol{x})$ is positive – Revenue flow grows up in the direction of vector $\boldsymbol{x}$. The same relations increase Revenue-on-Assets density $B(t,\boldsymbol{x})$ in unit volume at point $(t,\boldsymbol{x})$ if Assets flow $\boldsymbol{PA}(t,\boldsymbol{x}) = \boldsymbol{v}(t,\boldsymbol{x})A(t,\boldsymbol{x})$in the direction of vector $\boldsymbol{x}$ is positive. Simply speaking let assume that additional flow of Assets at point $\boldsymbol{x}$ increase Revenue-on-Assets at this point and vise versa. These assumptions neglect time gaps between Investments into Assets at point $\boldsymbol{x}$ and Revenue-on-Assets received and other factors that may impact on Assets allocations at point $\boldsymbol{x}$ and Revenue-on-Assets to simplify relations between them. Let take equations (3.1) on Assets density $B(t,\boldsymbol{x})$ Revenue-on-Assets density $B(t,\boldsymbol{x})$ as

$$\frac{\partial A}{\partial t} + \nabla \cdot (\boldsymbol{v}A) = a\,\boldsymbol{x} \cdot \boldsymbol{PB}(t,\boldsymbol{x}) \;;\; \frac{\partial B}{\partial t} + \nabla \cdot (\boldsymbol{u}B) = b\,\boldsymbol{x} \cdot \boldsymbol{PA}(t,\boldsymbol{x}) \tag{5.3}$$

To identify Equations of Motion (3.2) on impulses $\boldsymbol{PA}(t,\boldsymbol{x})$ and $\boldsymbol{PB}(t,\boldsymbol{x})$ let assume that $\boldsymbol{Q_2}$ is proportional to *conjugate* impulse. In other words let assume that growth of Assets impulse $\boldsymbol{PA}(t,\boldsymbol{x})$ is proportional to Revenue-on-Assets impulse $\boldsymbol{PB}(t,\boldsymbol{x})$ and vise versa. Let take Equations of Motion (3.2) as:

$$\frac{\partial \boldsymbol{PA}}{\partial t} + \nabla \cdot (\boldsymbol{v}\,\boldsymbol{PA}) = c\,\boldsymbol{PB}(t,\boldsymbol{x}) \;;\; \frac{\partial \boldsymbol{PB}}{\partial t} + \nabla \cdot (\boldsymbol{u}\,\boldsymbol{PB}) = d\,\boldsymbol{PA}(t,\boldsymbol{x}) \tag{5.4}$$

Economic meaning of (5.4) is follows. E-particles (economic agents) of entire economics fill certain domain on $n$-dimensional e-space. Due to current practice of international rating agencies [21-23] agent's risk ratings take minimum or most secure and maximum or most risky values. Thus let assume that coordinates $\boldsymbol{x}=(x_1,\dots x_n)$ of economic domain on $n$-dimensional e-space are reduced by relations

$$0 < x_i < X_i\,, i = 1, \dots n \tag{5.5}$$

As we mentioned above, we assume that small $x_i \ll 1$ correspond to secure area and

$$X_i - x_i \ll 1\,, i = 1, \dots n$$

to most risky area. Thus flow of Assets in such a bounded domain can't grow up forever. Assets flow can move from low risks and secure area of economic domain to high risks area and that increase Assets risks. Then Assets tide goes back from high risks to low risk area. Thus tides of Assets that is described by Assets impulse ***PA(t,x)***

$$\boldsymbol{PA}(t,\boldsymbol{x}) = \boldsymbol{v}(t,\boldsymbol{x})A(t,\boldsymbol{x})$$

follow certain fluctuations with frequency $\omega$ on economic domain of e-space. Let assume that

$$\omega^2 = -cd > 0 \qquad (5.6)$$

As we show below equations (5.4) describe simple fluctuations of aggregate Assets impulse ***PA(t)*** and aggregate Revenue-on-Assets impulse ***PB(t)***. Let take model equations (5.3; 5.4) to study aggregate fluctuations of Assets *A(t)* and Revenue-on-Assets *B(t).*

$$A(t) = \int d\boldsymbol{x} \;\; A(t,\boldsymbol{x}) \;\; ; \quad B(t) = \int d\boldsymbol{x} \;\; B(t,\boldsymbol{x}) \qquad (5.7)$$

## *3.2. Aggregate fluctuations of Assets A(t)*

Equations (5.3; 5.4) on Assets *A(t,x)* and Revenue-on-Assets *B(t,x)* densities allow derive ordinary differential equations on aggregate Assets *A(t)* and Revenue-on-Assets *B(t)* (Appendix A). Here we briefly argue main results.

Fluctuations of aggregate Assets *A(t)* depend on hidden aggregate variables *XPA(t)* and *YPB(t)* defined by (A.3); *EA(t)* (A.6) and *EB(t)* (A.7) (see Appendix A). Equations (A.5) on variables *XPA(t)* and *YPB(t)* (A.3) can be derived from Equations of Motion (5.4). *XPA(t)* is defined by (A.3) scalar product of vector $\boldsymbol{x}$ and impulse ***PA(t,x)***. Due to (5.6) variables *XPA(t)* follow oscillations with frequency $\omega$. Aggregate variable *EA(t)* (A.6) is proportional to Assets *A(t,x)* and squares of velocities $v^2(t,\boldsymbol{x})$ and can be treated *alike to* Assets energy. Same concern Revenue-on-Assets energy *EB(t)* that is *alike to* energy of hydrodynamic flow. To derive equations on aggregate Assets we should complement Continuity Equations (5.3) and Equations of Motion (5.4) with additional equations (A.8) on Assets energy *EA(t,x)* and Revenue-on-Assets energy *EB(t,x).* Equations (A.8) assume that aggregate Assets energy *EA(t)* and Revenue-on-Assets energy *EB(t)* grow up in time as exponent. (A.10) Fluctuations of Assets impulses ***PA(t)*** and ***PB(t)*** (B.5) with frequency $\omega$ (5.6) lead to fluctuations of *XPA(t)* and *YPB(t)* (A.5 ) with same frequency. Exponential growth of Assets energy *EA(t)* and *EB(t)* (A.10) and fluctuations of *XPA(t)* and *YPB(t)* (A.5) cause fluctuations of aggregate assets *A(t)* with frequency $\omega$ near exponential growth trend $exp(\gamma_e t)$. Aggregate Assets *A(t)* equals:

$$A(t) = A(1) + A(2)\cos\omega t + A(3)\sin\omega t + A(4)\,exp(\gamma_e t) + A(5)\,exp(-\gamma_e t) \qquad (5.8)$$

with constants *A(j), j=1,..5* determined by initial values (A.11-A.13).

### *3.3. Fluctuations of Assets mean risk X(t)*

Equations (5.3; 5.4) allow describe dynamics and oscillations of mean risks ***X(t)*** and mean square risks $\overline{X^2}(t)$ defined by Assets and Revenue-on-Assets densities. As we mentioned above (4.3 - 4.5) extensive (additive) aggregate economic or financial densities like Assets and Investment, Credits and Loans, Demand and Supply and etc., may define corresponding probability distributions like (4.3). Statistical moments of these distributions define mean risks ***X(t)***, mean square risks $\overline{X^2}(t)$ and so on and can be used as properties of economic evolution phases. Business cycles as fluctuations of aggregate variables induce oscillations of corresponding mean risks, mean squares and etc. It is obvious that different aggregate variables define different mean risks functions, and their oscillations may have different frequencies. Equations (5.3; 5.4) on Assets and Revenue-on-Assets densities allow derive equations on mean risks ***X(t)*** and mean squares risks $\overline{X^2}(t)$ defined by Assets and Revenue-on-Assets densities (Appendices B and C) but that requires introduction of additional aggregate variables and derivation corresponding additional hydrodynamic-like equations. Appendix B presents derivation of ordinary differential equations that describe fluctuations of mean Assets risks ***X(t)*** and Appendix C presents derivation of equations that describe $\overline{X^2}(t)$.

Assets mean risk ***X(t)*** and Revenue-on-Assets mean risk ***Y(t)*** depend upon aggregate Assets *A(t)* and aggregate Revenue-on-Assets *B(t)* and on new variables defined in Appendix B as aggregate Assets impulses ***PA(t)*** and ***PB(t)*** (B.3); aggregate factors ***XP(t)*** and ***YP(t)*** (equations B.4); aggregate Assets energy risks ***XE(t)*** and aggregate Revenue-on-Assets energy risks ***YE(t)*** (equations B.8); on aggregate Assets energy impulses ***PEA(t)*** and aggregate Revenue-on-Assets energy impulses ***PEB(t)*** (equations B.14). Aggregate Assets *A(t)* and Assets mean risk ***X(t)*** (B.19) takes form:

$$\boldsymbol{X}(t)A(t) = \boldsymbol{XA}(1) + \boldsymbol{XA}(2)\sin\omega t + \boldsymbol{XA}(3)\cos\omega t + \boldsymbol{XA}(4)\cos\omega_{pe}t + \boldsymbol{XA}(5)\ \sin\omega_{pe}t$$
$$+ \boldsymbol{XA}(6)\cos\omega_v t + \boldsymbol{XA}(7)\ \sin\omega_v t + \boldsymbol{XA}(8)\exp(\gamma_e t) + \boldsymbol{XA}(9)\exp(-\gamma_e t)$$

Constants $\boldsymbol{XA}(j), j = 1,..9$ are defined by initial values of (B.18.1-B.18.6). $\boldsymbol{X}(t)A(t)$ has same exponential increment $\gamma_e$ as aggregate assets *A(t)*. Hence Assets mean risk ***X(t)*** don't grow up but follow only time oscillations with frequencies $\omega$ (A.11) as aggregate Assets *A(t)* and has additional frequencies $\omega_{pe}$ (B.15) induced by oscillations of "energy impulses" ***PEA(t)*** and ***PEB(t)*** (B.13) and frequencies $\omega_v$ induced by oscillations of factors ***VXPA*** and ***UYPB*** (B.17.3). Assets mean risk ***X(t)*** describes evolution of Assets distribution over e-space and its fluctuations reflect main drivers of Assets dynamics.

### *3.4. Fluctuations of Assets mean square risk* $\overline{X^2}(t)$

Assets mean square risk $\overline{X^2}(t)$ and Revenue-on-Assets mean square risk $\overline{Y^2}(t)$ describe distribution of Assets and Revenue-on-Assets on e-space (Appendix C). Assets mean square risk $\overline{X^2}(t)$ defines Assets dispersion ${\sigma_A}^2(x)$

$$ {\sigma_A}^2(x) = \overline{X^2}(t) - \boldsymbol{X}^2(t) $$

Growth of Assets dispersion ${\sigma_A}^2(x)$ reflects growth on Assets risk uncertainty and decrease of ${\sigma_A}^2(x)$ describes clustering of Assets near its mean risk. Assets mean squares risk $\overline{X^2}(t)A(t)$ (Appendix C) depend on new factors different from those define aggregate Assets *A(t)* and mean risks $\boldsymbol{X}(t)$.

$$ \overline{X^2}(t)A(t) = \overline{X^2}A(0) + \overline{X^2}A(1)\sin\omega t + \overline{X^2}A(2)\cos\omega t + \overline{X^2}A(3)\sin\omega_{pe}t $$
$$ + \overline{X^2}A(4)\exp\gamma_e t + \overline{X^2}A(5)\ \exp-\gamma_e t + \overline{X^2}A(6)\ \exp\gamma_{vu}t $$
$$ + \overline{X^2}A(7)\ \exp-\gamma_{vu}t + \overline{X^2}A(8)\ \exp\gamma_{xv}t + \overline{X^2}A(9)\ \exp-\gamma_{xv}t $$

with constants $\overline{X^2}A(j)$, *j=0,..9* defined by initial values of variables (A.3; A.6; A.7; C.6; C.7.2; C.10.1; C12.2) for *t=0.* Factor $\overline{X^2}(t)A(t)$ fluctuates with frequencies $\omega$ and $\omega_{pe}$ (5.4; B.13) and has exponential growth increments $\gamma_e$ ,$\gamma_{vu}$,$\gamma_{xv}$ (A.9; C.12.3; C.15). Aggregate Assets *A(t)* has exponential increment $\gamma_e$ (A.9; A.14). As $\overline{X^2}(t)$ can't grow up as exponent hence it is required that $\gamma_e > \gamma_{vu}$ ; $\gamma_e > \gamma_{xv}$.

## 4. Conclusions

We present a general approach for modeling aggregate fluctuations for any system of economic and financial variables. Distributions of economic agents by their risk ratings as coordinates on economic space allow describe economic and financial states and processes by hydrodynamic-like equations. Interactions between four-five or more *conjugate* economic or financial variables can be modeled by equations (3.1; 3.2). Such representation uncovers hidden complexity of financial and economic processes and gives opportunity to apply a wide range of mathematical physics methods to financial and economic modeling. Vital distinctions between economic and physical systems prohibit any direct application of known results. Huge amount of variables that impact economic and financial evolution makes such description extremely difficult.

Economic nature of business cycle fluctuations of aggregate variables is determined by repeated motion of economic agents and their variables from low risk to high-risk area on economic space and back. Fluctuations of agents between low and high-risk areas on economic space induce oscillations of aggregate fluctuations with various frequencies. We model such tides of economic and financial variables from low risks to high risks on economic space and back by simple equations (5.4; B.13; B.17.3) that induce harmonique oscillations with frequencies

$$\omega^2 = -cd > 0 \quad ; \omega_{pe}{}^2 = -c_{ep}\, d_{ep} > 0 \; ; \; \omega_v^2 = -c_v d_v > 0$$

Interactions between numerous economic and financial variables cause complex structure for fluctuating frequencies and can explain variety scales of aggregate fluctuations. Growth of hidden variables induces exponential growth of aggregated economic factors with increments (A.9; C.12.3; C.15):

$$\gamma_e{}^2 = c_e d_e > 0 \; ; \gamma^2{}_{vu} = c_u d_v > 0 \; ; \gamma_{xv}{}^2 = c_{xv} d_{xv} > 0$$

We present simplest model of mutual dependence between two economic variables – Assets and Revenue-on-Assets to demonstrate advantages of our approach and to present simple derivation of equations that describe aggregate fluctuations of these variables. Even such simple model shows certain complexity and indicates that attempts to describe mutual relations between four-five and more variables will lead to modeling solutions for huge system of hydrodynamic-like equations and will require much more efforts and computer programs. The most interesting issue concerns structure of variables that define aggregate fluctuations of Assets. Nevertheless we model interaction between two variables - Assets and Revenue-on-Assets, system of hydrodynamic-like equations describe Assets density by Continuity equation (5.3), Assets impulses by Equation of Motion (5.4) and additional unexpected variable alike to Assets energy (A.6) and Revenue-on-Assets energy (A.7) that are described by equation (A.8). Naturally there are no substantive parallels between energy of hydrodynamic flow and Assets energy (A.8). Anyway, usage of this hidden variable extends parallels between economics and hydrodynamics.

Let underline – we model interaction between Assets *A(t,**x**)* and Revenue-on-Assets *B(t,**x**)* on economic space and that allow derive equations on aggregate Assets *A(t)* and aggregate Revenue-on-Assets *B(t)*. Aggregate Assets fluctuations don't depend on aggregate Revenue-on-Assets, but on variables *alike to* aggregate Revenue-on-Assets energy and other (A.7; A.3). Thus it seems unbelievable to describe business cycles by models that take into account only direct dependence between time-series of aggregate variables as in equilibrium-based models of business cycles.

Economic evolution is a very complex processes and description of evolution phases requires much more then description of aggregate fluctuations. Hydrodynamic-like equations allow describe additional parameters of economic evolution phases that can be noted as mean risks, mean square risks, dispersions and etc. Indeed, most economic densities on e-space are positive and thus can be treated as certain probability distributions on e-space. For example Assets distribution of e-space (4.2-4.5) can define Assets mean risks $\boldsymbol{X(t)}$ (6) and Assets mean square $\overline{X^2}(t)$ (7). Each economic and financial variable can define its own mean risks and mean squares and these functions describe risk state and risk fluctuations of particular variable on economic space. Business cycles as aggregate fluctuations are accompanied by corresponding fluctuations of risk statistical moments of corresponding variables like mean risk or square risk. These parameters give additional information about uncertainty of economic variables and might be useful for economic and financial managing and policy-making. Complexities of relations (Appendix B and C) that determine dynamics of risk statistical moments leave few chances to derive similar relations by "mainstream" models.

Current econometric data don't have sufficient info that is required for economic space modeling, definition of Assets $A(t,\boldsymbol{x})$ and Revenue-on-Assets $B(t,\boldsymbol{x})$ distributions and their modeling by hydrodynamic-like equations. Thus our theoretical description of business cycle aggregate fluctuations can't be verified by existing econometric data. Nevertheless we propose that there are no inscrutable and unsolvable problems that might prevent enhancement of risk ratings methodologies and development of econometric data to support economic modeling on risk ratings economic space. Proposed development will help better understand and respond economic and financial processes and improve policymaking.

## Appendix A

## Aggregate Assets *A(t)* and Revenue-on-Assets *B(t)* Equations

To derive equations on aggregate Assets *A(t)* let start with (5.3; 5.6):

$$\frac{d}{dt}A(t) = \int dx\ \frac{\partial}{\partial t}A(t,\boldsymbol{x}) = -\int dx\ \nabla\cdot(\boldsymbol{v}A) + a\ \int dx\,\boldsymbol{x}\ \cdot\boldsymbol{PB} \qquad (A.1)$$

Due to (5.5) economic domain on e-space [17,18] is reduced by most secure and by most risky e-particles (economic agents). Hence integral of divergence over e-space equals integral of flux through surface and that equals zero as no economic or financial fluxes exist far from boundaries (5.6). Thus first integral in the right hand of (A.1) equals zero. Equations on *A(t)* and *B(t)* take form:

$$\frac{d}{dt}A(t) = \ a\,YPB(t)\ ;\ \frac{d}{dt}B(t) = \ b\,XPA(t) \qquad (A.2)$$

$$XPA(t) = \int dx\,\boldsymbol{x}\ \cdot\boldsymbol{PA(t,x)}\ ;\ \ YPB(t) = \int dx\,\boldsymbol{x}\ \cdot\boldsymbol{PB(t,x)} \qquad (A.3)$$

To derive equations on *XPA(t)* и *YPB(t)* let use equation (5.4):

$$\frac{d}{dt}XPA(t) = \int dx\,\boldsymbol{x}\ \cdot\frac{\partial}{\partial t}\boldsymbol{PA}(t,\boldsymbol{x}) = -\int dx\,\boldsymbol{x}\ \cdot\nabla\cdot(\boldsymbol{v}\,\boldsymbol{PA}) + c\,YPB(t) \qquad (A.4)$$

Let take first integral in the right hand by parts and take into account that integral by divergence equals zero:

$$-\int dx\,\boldsymbol{x}\ \cdot\nabla\cdot(\boldsymbol{v}\,\boldsymbol{PA}) = \int dx\ \ A(t,\boldsymbol{x})v^2(t,\boldsymbol{x})$$

$$\frac{d}{dt}XPA(t) = EA(t) + c\,YPB(t)\ \ ;\ \ \ \frac{d}{dt}YPB(t) = EB(t) + d\,XPA(t) \qquad (A.5)$$

$$EA(t) = v^2(t)A(t) = \int dx\ \ A(t,\boldsymbol{x})v^2(t,\boldsymbol{x}) = \int dx\ \ A(t,\boldsymbol{x})\textstyle\sum_i {v_i}^2 \qquad (A.6)$$

$$EB(t) = u^2(t)B(t) = \int dx\ \ B(t,\boldsymbol{x})u^2(t,\boldsymbol{x}) = \int dx\ \ B(t,\boldsymbol{x})\textstyle\sum_i {u_i}^2 \qquad (A.7)$$

It is amazing that equations on aggregate Assets and Revenue-on-Assets require factors (A.6; A.7) that are *alike to* Assets energy *EA(t)* and Revenue-on-Assets energy *EB(t).* These factors enhance parallels between economic modeling and hydrodynamics. It is obvious that Assets *energy EA(t)* has only formal resemblance with physical energy used in hydrodynamics but as well it underlines very important issue: adequate modeling of aggregate macroeconomic variables can't be based on usual economic variables only but require wide range of derivative factors like (A.3; A.6; A.7). To describe aggregate Assets *A(t)* in a closed form let take hydrodynamic-like equations on Assets energy density *EA(t,$\boldsymbol{x}$)* and Revenue-on-Assets energy density *EB(t,$\boldsymbol{x}$)* as:

$$\frac{\partial EA}{\partial t} + \nabla\cdot(\boldsymbol{v}EA) = c_e\,EB(t,\boldsymbol{x})\ ;\ \ \frac{\partial EB}{\partial t} + \nabla\cdot(\boldsymbol{u}EB) = d_e\,EA(t,\boldsymbol{x}) \qquad (A.8)$$

Economic meaning of (A.8) is follows. Flows of Assets $A(t,\boldsymbol{x})$ from low risk area to high risk area and back are accompanied with growth of "Assets energy" proportional to growth of aggregate Assets A(t) and square of velocity $v^2(t)$. The same happens to flows of Revenue-on-Assets energy $EB(t,\boldsymbol{x})$. Assumptions for right side of equations (A.8) for

$$\gamma_e{}^2 = c_e d_e > 0 \tag{A.9}$$

mean that aggregate Assets energy $EA(t)$ and aggregate Revenue-on-Assets energy $EB(t)$ grow up as exponent and (A.8; A.9) give equations on aggregate energies $EA(t)$ and $EA(t)$:

$$\frac{d}{dt}EA(t) = c_e\, EB(t) \quad ; \quad \frac{d}{dt}EB(t) = d_e\, EA(t) \tag{A.10}$$

That allows close sequence of equations on aggregate Assets. Equations (A.2; A.5; A10) form a system of equations that describe aggregate Assets $A(t)$ and Revenue-on-Assets $B(t)$. Initial values of variables are defined as:

$$A(0) = \int dx\, A(0,\boldsymbol{x})\,; \quad B(0) = \int dx\, B(0,\boldsymbol{x})\,; \tag{A.11}$$

$$XPA(0) = \int dx\, \boldsymbol{x} \cdot \boldsymbol{v}(0,\boldsymbol{x})A(0,\boldsymbol{x})\,; \quad YPB(0) = \int dx\, \boldsymbol{x} \cdot \boldsymbol{u}(0,\boldsymbol{x})B(0,\boldsymbol{x}) \tag{A.12}$$

$$EA(0) = \int dx\, v^2(0,\boldsymbol{x})A(0,\boldsymbol{x})\,; \quad EB(0) = \int dx\, u^2(0,\boldsymbol{x})B(0,\boldsymbol{x}) \tag{A.13}$$

It is easy to show that initial values (A.11-A.13) define solution for aggregate Assets $A(t)$ as:

$$A(t) = A(1) + A(2)\cos\omega t + A(3)\sin\omega t + A(4)\,exp(\gamma_e t) + A(5)\,exp(-\gamma_e t) \tag{A.14}$$

*A(j)* for *j=1,..5* are constants that are defined by (A.11-A.13) and simple solutions of Equations (A.2; A.5; A10) and we omit here exact derivation for brevity. Aggregate Assets *A(t)* and Revenue-on-Assets *B(t)* takes form similar to (A.14) with its own constants *B(j)*, *j=1,..5*.

$$B(t) = B(1) + B(2)\cos\omega t + B(3)\sin\omega t + B(4)\,exp(\gamma_e t) + B(5)\,exp(-\gamma_e t) \tag{A.15}$$

Main conclusion: simple assumptions on exponential growth of aggregate Assets and Revenue-on-Assets energy with increment $\gamma_e$ and oscillations of *XAP(t)* and *YAP(t)* with frequency $\omega$ induce fluctuations of aggregate Assets with frequency $\omega$ around exponential growth trend. Aggregate assets growth and fluctuations don't depend directly on conjugate variable Revenue-on-Assets, but on hidden variables *XAP(t)*, *YAP(t), EA(t), EB(t)* that are alike to hydrodynamic impulses and energies.

## Appendix B

### Assets and Revenue-on-Assets mean risks *X(t)* and *Y(t)*

Derivations of Assets mean risks follow same scheme as Appendix A. Due to equation (5.3):

$$\frac{d}{dt}\boldsymbol{X}(t)A(t) = \int dx\,\boldsymbol{x}\,\frac{\partial A}{\partial t} = -\int dx\,\boldsymbol{x}\,\nabla\cdot(\boldsymbol{v}A) + a\int dx\,\boldsymbol{x}\,(\,\boldsymbol{x}\,\cdot\boldsymbol{PB}) \quad \text{(B.1)}$$

Let take first integral in the right side by parts and take into account that integral by divergence over e-space equals zero:

$$-\int dx\,\boldsymbol{x}\,\nabla\cdot(\boldsymbol{v}A) = \int dx\,\boldsymbol{v}A(t,x) = \boldsymbol{PA}(t)$$

$$\frac{d}{dt}\boldsymbol{X}(t)A(t) = \boldsymbol{PA}(t) + a\,\boldsymbol{YP}(t)\,;\quad \frac{d}{dt}\boldsymbol{Y}(t)B(t) = \boldsymbol{PB}(t) + b\,\boldsymbol{XP}(t) \quad \text{(B.2)}$$

$$\boldsymbol{PA}(t) = \int dx\;\boldsymbol{PA}(t,x) = \int dx\,\boldsymbol{v}A(t,x)\;;\;\boldsymbol{PB}(t) = \int dx\;\boldsymbol{PB}(t,x) = \int dx\,\boldsymbol{u}B(t,x) \quad \text{(B.3)}$$

$$\boldsymbol{XP}(t) = \int dx\,\boldsymbol{x}\left(\,\boldsymbol{x}\,\cdot\boldsymbol{PA}\,(t,x)\right)\;;\quad \boldsymbol{YP}(t) = \int dx\,\boldsymbol{x}\left(\,\boldsymbol{x}\,\cdot\boldsymbol{PB}\,(t,x)\right) \quad \text{(B.4)}$$

Equations on aggregate Assets impulse ***PA(t)*** appears from (3.4)

$$\frac{d}{dt}\boldsymbol{PA}(t) = \int dx\,\frac{\partial \boldsymbol{PA}}{\partial t} = -\int dx\,\nabla\cdot(\boldsymbol{v}\,\boldsymbol{PA}) + c\int dx\;\;\boldsymbol{PB}$$

First integral in the right side equals zero and equations take form:

$$\frac{d}{dt}\boldsymbol{PA}(t) = c\;\;\boldsymbol{PB}(t)\;\;;\;\frac{d}{dt}\boldsymbol{PB}(t) = d\;\;\boldsymbol{PA}(t) \quad \text{(B.5)}$$

To derive equations on ***XP(t)*** and ***YP(t)*** let use equations (3.4):

$$\frac{d}{dt}\boldsymbol{XP}(t) = \int dx\,\boldsymbol{x}\left(\boldsymbol{x}\cdot\frac{\partial \boldsymbol{PA}}{\partial t}\right) = -\int dx\,\boldsymbol{x}\left(\boldsymbol{x}\cdot\nabla\cdot(\boldsymbol{v}\,\boldsymbol{PA})\right) + \;c\boldsymbol{YP}(t) \quad \text{(B.6.1)}$$

$$\frac{d}{dt}\boldsymbol{YP}(t) = -\int dx\,\boldsymbol{x}\left(\boldsymbol{x}\cdot\nabla\cdot(\boldsymbol{u}\,\boldsymbol{PB})\right) + \;d\,\boldsymbol{XP}(t) \quad \text{(B.6.2)}$$

We omit long but simple calculations of integral in (B.6.2) and present the result as:

$$\int dx\;\;\boldsymbol{x}\,(\boldsymbol{x}\;\cdot(\nabla\cdot\left(\boldsymbol{v}\,\boldsymbol{PA}(t,\boldsymbol{x})\right))) = -\int dx\;\;\boldsymbol{x}\,EA(t) - \int dx\,\boldsymbol{v}\,(\boldsymbol{x}\cdot\boldsymbol{PA}(t,\boldsymbol{x})) \quad \text{(B.7)}$$

(B.7) define new factors:

$$\boldsymbol{XE}(t) = \int dx\,\boldsymbol{x}\,EA(t,x)\;\;;\;\;\boldsymbol{YE}(t) = \int dx\,\boldsymbol{x}\,EB(t,x) \quad \text{(B.8.1)}$$

$$\boldsymbol{VXPA}(t) = \int dx\,\boldsymbol{v}(t,\boldsymbol{x})\left(\boldsymbol{x}\cdot\boldsymbol{PA}(t,\boldsymbol{x})\right) = \int dx\;\;\boldsymbol{VXPA}(t,\boldsymbol{x}) \quad \text{(B.8.2)}$$

$$\boldsymbol{UYPB}(t) = \int dx\,\boldsymbol{u}(t,\boldsymbol{x})\left(\boldsymbol{x}\cdot\boldsymbol{PB}(t,\boldsymbol{x})\right) = \int dx\;\;\boldsymbol{UYPB}(t,\boldsymbol{x}) \quad \text{(B.8.3)}$$

Equations (B.6.1; B.6.2) take form:

$$\frac{d}{dt}\boldsymbol{XP}(t) = \boldsymbol{XE}(t) + \boldsymbol{VXPA}(t) + \;c\,\boldsymbol{YP}(t) \quad \text{(B.9.1)}$$

$$\frac{d}{dt}\boldsymbol{YP}(t) = \boldsymbol{YE}(t) + \boldsymbol{UYPB}(t) + \;d\,\boldsymbol{XP}(t) \quad \text{(B.9.2)}$$

Equations (A.8) on Assets energy allow derive equations on $\boldsymbol{XE}(t)$ и $\boldsymbol{YE}(t)$:

$$\frac{d}{dt}\boldsymbol{XE}(t) = \int dx\,\boldsymbol{x}\,\frac{\partial EA}{\partial t} = -\int dx\,\boldsymbol{x}\,\nabla\cdot(\boldsymbol{v}EA) + c_e\boldsymbol{YE}(t) \tag{B.10}$$

Alike to equations (A.5) or (B.6.1) first integral in the right side can be transformed to

$$-\int dx\,\boldsymbol{x}\,\nabla\cdot(\boldsymbol{v}EA) = \int dx\;\boldsymbol{v}\,EA(t,\boldsymbol{x}) = \int dx\;\boldsymbol{PEA}(t,\boldsymbol{x})$$

factor in the right side is similar to energy impulse

$$\boldsymbol{PEA}(t,\boldsymbol{x}) = \boldsymbol{v}(t,\boldsymbol{x})\,EA(t,\boldsymbol{x}) \;\;;\;\; \boldsymbol{PEB}(t,\boldsymbol{x}) = \boldsymbol{u}(t,\boldsymbol{x})\,EB(t,\boldsymbol{x}) \tag{B.11}$$

Let complement initial equations (3.3; 3.4; A.8) with hydrodynamic-like equations on energy impulse as:

$$\frac{\partial \boldsymbol{PEA}}{\partial t} + \nabla\cdot(\boldsymbol{vPEA}) = c_{pe}\,\boldsymbol{PEB}(t,\boldsymbol{x})\,;\; \frac{\partial \boldsymbol{PEB}}{\partial t} + \nabla\cdot(\boldsymbol{u}\,\boldsymbol{PEB}) = d_{pe}\,\boldsymbol{PEA}(t,\boldsymbol{x}) \tag{B.12}$$

Here we assume that energy impulses ***PEA(t,x)*** and ***PEB(t,x)*** fluctuate with frequency $\omega_{pe}$ due to same reasons as fluctuations of Assets impulse ***PA(t,x)*** with frequency $\omega$ (5.6):

$$\omega_{pe}{}^2 = -c_{ep}\,d_{ep} > 0 \tag{B.13}$$

Equations (B.12) allow derive equations on aggregate energy impulse ***PEA(t)*** and ***PEB(t)*** as:

$$\boldsymbol{PEA}(t) = \int dx\,\boldsymbol{PEA}(t,\boldsymbol{x}) \;\;;\; \boldsymbol{PEB}(t) = \int dx\,\boldsymbol{PEB}(t,\boldsymbol{x}) \tag{B.14}$$

$$\frac{d}{dt}\boldsymbol{PEA}(t) = c_{pe}\boldsymbol{PEB}(t) \;\;;\;\; \frac{d}{dt}\boldsymbol{PEB}(t) = d_{pe}\boldsymbol{PEA}(t) \tag{B.15}$$

Equations (B.15) describe oscillations of ***PEA(t)*** and ***PEB(t)*** with frequency $\omega_{pe}$. Thus equations (B.10) take form:

$$\frac{d}{dt}\boldsymbol{XE}(t) = \boldsymbol{PEA}(t) + c_e\boldsymbol{YE}(t,x) \;\;;\;\; \frac{d}{dt}\boldsymbol{YE}(t) = \boldsymbol{PEB}(t) + d_e\boldsymbol{XE}(t,x) \tag{B.16}$$

To define equations (B.9.2; B.9.3) on $\boldsymbol{XP}(t)$, $\boldsymbol{YP}(t)$ let describe factors ***VXPA*** and ***UYPB*** (B.8.2; B.8.3). We propose that these factors follow similar fluctuations as impulses ***PA(t,x)*** and ***PB(t,x)*** that are described by (5.4; B.5) or factors ***PEA*** and ***PEB*** (B.12-B.15). Such fluctuations are induced my motion of e-particles from low risk to high-risk area and back on e-space. Let take equations on $\boldsymbol{VXPA}(t,\boldsymbol{x})$ and $\boldsymbol{UYPB}(t,\boldsymbol{x})$ as:

$$\frac{\partial \boldsymbol{VXPA}}{\partial t} + \nabla\cdot(\boldsymbol{v}\,\boldsymbol{VXPA}) = c_v\boldsymbol{UYPB}(t,\boldsymbol{x})\,;\; \frac{\partial \boldsymbol{UYPB}}{\partial t} + \nabla\cdot(\boldsymbol{u}\,\boldsymbol{UYPB}) = d_v\boldsymbol{VXPA}(t,\boldsymbol{x}) \tag{B.17.1}$$

$$\frac{d}{dt}\boldsymbol{VXPA}(t) = c_v\;\boldsymbol{UYPB}(t) \;\;;\;\; \frac{d}{dt}\boldsymbol{UYPB}(t) = d_v\;\boldsymbol{VXPA}(t) \tag{B.17.2}$$

$$\omega_v^2 = -c_v d_v > 0 \tag{B.17.3}$$

Then equations on ***XP(t)*** and ***YP(t)*** (B.9.2; B.9.3) are determined. Hence we obtain system of equations (B.2; B.5; B.9.1; B.9.2; B.15; B.16; B.17.2) on Assets mean risk ***X(t)*** and on Revenue-on-Assets mean risk ***Y(t)***. Let define initial values as:

$$\boldsymbol{X}(0)A(0) = \int dx\,\boldsymbol{x}\,A(0,\boldsymbol{x}) \;\;;\;\; \boldsymbol{Y}(0)B(0) = \int dx\,\boldsymbol{x}\,B(0,\boldsymbol{x}) \tag{B.18.1}$$

$$\boldsymbol{PA}(0) = \int dx\,\boldsymbol{v}(0,\boldsymbol{x})\,A(0,\boldsymbol{x}) \;\;;\;\; \boldsymbol{PB}(0) = \int dx\,\boldsymbol{u}(0,\boldsymbol{x})\,B(0,\boldsymbol{x}) \tag{B.18.2}$$

$$\boldsymbol{XP}(0) = \int dx\, \boldsymbol{x}\ \boldsymbol{x} \cdot \boldsymbol{v}(0,\boldsymbol{x}) A\,(0,\boldsymbol{x}) \ \ ; \ \ \boldsymbol{YP}(0) = \int dx\, \boldsymbol{x}\ \boldsymbol{x} \cdot \boldsymbol{u}(0,\boldsymbol{x}) B\,(0,\boldsymbol{x}) \qquad \text{(B.18.3)}$$

$$\boldsymbol{PEA}(0) = \int dx\, \boldsymbol{v}(0,\boldsymbol{x}) EA(0,\boldsymbol{x}) \ ; \ \boldsymbol{PEB}(0) = \int dx\, \boldsymbol{u}(0,\boldsymbol{x}) EB(0,\boldsymbol{x}) \qquad \text{(B.18.4)}$$

$$\boldsymbol{XE}(0) = \int dx\, \boldsymbol{x}\, v^2(0,x)\, A(0,x); \ \ \boldsymbol{YE}(t) = \int dx\, \boldsymbol{x}\, u^2(0,x)\, B(0,x) \qquad \text{(B.18.5)}$$

$$\boldsymbol{VXPA}(0) = \int dx\ \boldsymbol{v}(0,\boldsymbol{x})\big(\boldsymbol{x} \cdot \boldsymbol{PA}(0,\boldsymbol{x})\big)\, ; \boldsymbol{UYPB}(0) = \int dx\ \boldsymbol{u}(0,\boldsymbol{x})\big(\boldsymbol{x} \cdot \boldsymbol{PB}(0,\boldsymbol{x})\big) \ \text{(B.18.6)}$$

We omit long but simple calculations that allow derive solution as

$$\boldsymbol{X}(t)A(t) = \boldsymbol{XA}(1) + \boldsymbol{XA}(2) \sin \omega t + \boldsymbol{XA}(3) \cos \omega t + \boldsymbol{XA}(4) \cos \omega_{pe} t + \boldsymbol{XA}(5)\ \sin \omega_{pe} t +$$
$$\boldsymbol{XA}(6) \cos \omega_v t + \boldsymbol{XA}(7)\ \sin \omega_v t + \boldsymbol{XA}(8)\, exp(\gamma_e t) + \boldsymbol{XA}(9)\, exp(-\gamma_e t) \qquad \text{(B.19)}$$

with constants *XA(j), j=1,..9* determined by initial values (B.18.1-6). Aggregate Assets *A(t)* (A.14) has same exponential trend with increment $\gamma_e$ thus mean risks ***X(t)*** don't grow up but follow fluctuations with frequencies $\omega$ and $\omega_{pe}$ and $\omega_v$ . Revenue-on-Assets mean risk ***Y(t)*** is determined by relations similar to (B.19) and (A.15).

## Appendix C

# Assets Mean Squares Risks $\overline{X^2}(t)$ and Dispersions

Let take Assets and Revenue-on-assets mean square risks $\overline{X^2}(t)$ and $\overline{Y^2}(t)$ as

$$\overline{X^2}(t)A(t) = \int dx \;\; x^2 \; A(t,x) \;\; ; \;\; \overline{Y^2}(t)B(t) = \int dx \;\; x^2 \; B(t,x) \qquad \text{(C.1)}$$

Then Assets dispersion

$$\sigma_A^2(t) = \overline{X^2}(t) - X^2(t) \qquad \text{(C2)}$$

Assets dispersion describes width of Assets distribution on e-space. Growth of (C.2) indicates increase of spreading Assets over risks on e-space. Small Assets dispersion means that most Assets of macroeconomics are clustered near mean Assets risks ***X(t)***. To describe Assets dispersion let derive equations on Assets mean squares risks. Let use (C.1) and (5.3):

$$\frac{d}{dt}\,\overline{X^2}(t)A(t) = -\int dx \; \sum_i x_i^2 \;\; \nabla \cdot (vA) + \; a\int dx \; \sum_i x_i^2 \;\; \boldsymbol{x} \; \cdot \boldsymbol{PB} \qquad \text{(C3)}$$

For each *i* first integral in the right side equals

$$\int dx \;\; x_i^2 \; \nabla_j \cdot (vA) = \int dx_i \, x_i^2 \int dx_{k\neq i} \, \nabla_j \cdot (vA) + \int dx \, x_i^2 \nabla_i \cdot (vA)$$

As we mentioned above integral over divergence equals zero and

$$\int dx \, x_i^2 \nabla_i \cdot (vA) = \int dx_{k\neq i} \int dx_i \, x_i^2 \nabla_i \cdot (v_i A) = -2\int dx \, x_i \cdot v_i A$$

Thus obtain

$$\int dx \, x^2 \nabla \cdot (vA) = -2\int dx \;\; \boldsymbol{x} \; \cdot \boldsymbol{v}(t,\boldsymbol{x}) \; A(t,\boldsymbol{x}) = -2 \, XPA(t) \qquad \text{(C4)}$$

(A.3; A.5) define *XPA(t)* and *YPA(t)*. Equations (C.3) take form:

$$\frac{d}{dt}\,\overline{X^2}(t)A(t) = 2\,XPA + \; aYPBY^2(t) \;\; ; \;\; \frac{d}{dt}\,\overline{Y^2}(t)B(t) = 2\,YPB + \; bXPAX^2(t) \qquad \text{(C.5)}$$

$$XPAX^2(t) = \int dx \;\; x^2 \;\; \boldsymbol{x} \; \cdot \boldsymbol{PA}(t,\boldsymbol{x}) \;\; ; \; YPBY^2(t) = \int dx \;\; x^2 \;\; \boldsymbol{x} \; \cdot \boldsymbol{PB}(t,\boldsymbol{x}) \qquad \text{(C.6)}$$

To derive equations on $XPAX^2(t)$ and $YPBY^2(t)$ let use (5.4):

$$\frac{d}{dt}XPAX^2(t) = -\int dx \, x^2 \;\; \boldsymbol{x} \; \cdot (\nabla \cdot (\boldsymbol{v}\,\boldsymbol{PA})) + c\,YPBY^2(t) \qquad \text{(C.7.1)}$$

We omit long but simple evaluation of integral in (C.7.1) and present the result as:

$$-\int dx \, x^2 \;\; \boldsymbol{x} \; \cdot (\nabla \cdot (\boldsymbol{v}\,\boldsymbol{PA})) = \int dx \;\; x^2 \; EA(t,\boldsymbol{x}) + 2\int dx \;\; (\boldsymbol{x} \cdot \boldsymbol{v})^{\boldsymbol{2}} A$$

$$X^2EA(t) = \int dx \;\; x^2 EA(t,\boldsymbol{x}) \;\;\; ; \;\; Y^2EB(t) = \int dx \;\; x^2 EB(t,\boldsymbol{x}) \qquad \text{(C.7.2)}$$

$$XVA(t) = \int dx \;\; (\boldsymbol{x} \cdot \boldsymbol{v}(t,\boldsymbol{x}))^{\boldsymbol{2}} A(t,\boldsymbol{x}) = \int dx \;\; XVA(t,\boldsymbol{x}) \qquad \text{(C.7.3)}$$

$$YUB(t) = \int dx \;\; (\boldsymbol{x} \cdot \boldsymbol{u}(t,\boldsymbol{x}))^{\boldsymbol{2}} B(t,\boldsymbol{x}) = \int dx \;\; YUB(t,\boldsymbol{x}) \qquad \text{(C.7.4)}$$

Thus equation (C.7.1) takes form:

$$\frac{d}{dt}XPAX^2(t) = X^2EA(t) + 2XVA(t) + c\,YPBY^2(t) \tag{C.8.1}$$

$$\frac{d}{dt}YPBY^2(t) = Y^2EB(t) + 2YUA(t) + d\,XPAX^2(t) \tag{C.8.2}$$

To derive equations on $X^2EA(t)$ and $Y^2EB(t)$ let use equations (A.8):

$$\frac{d}{dt}X^2EA(t) = -\int dx\;x^2\nabla\cdot(\boldsymbol{v}EA) + c_e\,Y^2EB(t)$$

$$\int dx\sum x_i^2\;\;\nabla\cdot(\boldsymbol{v}EA) = \int dx\;\sum x_i^2\;\;\nabla_{j\neq i}\cdot\left(v_j\,EA\right) + \int dx\sum x_i^2\;\;\nabla_i\cdot(v_i\,EA)$$

First integral by divergence equals zero and second integral by parts gives:

$$\int dx\sum x_i^2\;\;\nabla_i\cdot(v_i\,EA) = -2\;\int dx\;\;\boldsymbol{x}\cdot\boldsymbol{v}\,EA(t,x) = -2\int dx\;\;\boldsymbol{x}\cdot\boldsymbol{PEA}(t,x)$$

Thus obtain equations on $X^2EA(t)$ and $Y^2EB(t)$:

$$\frac{d}{dt}X^2EA(t) = 2XPEA(t) + c_eY^2EB(t)\,;\;\frac{d}{dt}Y^2EB(t) = 2YPEB(t) + d_eX^2EA(t) \tag{C.9}$$

$$XPEA(t) = \int dx\;\;\boldsymbol{x}\cdot\boldsymbol{PEA}(t,x)\;\;;\quad YPEB(t) = \int dx\;\;\boldsymbol{x}\cdot\boldsymbol{PEB}(t,x) \tag{C.10.1}$$

To derive equations on *XPEA(t)* and *YPEB(t)* let use equations (B.12) on energy impulse $\boldsymbol{PEA}(t,\boldsymbol{x})$ and $\boldsymbol{PEB}(t,\boldsymbol{x})$. Equations on *XPEA(t)* and *YPEB(t)* take form:

$$\frac{d}{dt}XPEA(t) = -\int dx\;\;\boldsymbol{x}\cdot\;\nabla\cdot(\boldsymbol{vPEA}) + c_{pe}\,YPEB(t) \tag{C.10.2}$$

$$\int dx\;\;\boldsymbol{x}\cdot\;\nabla\cdot(\boldsymbol{vPEA}) = \int dx\;\;x_j\;\nabla_{i\neq j}\;v_iv_jEA(t,\boldsymbol{x}) + \int dx\;\;x_j\;\nabla_jv_j^2\;EA(t,\boldsymbol{x})$$

First integral by divergence equals zero and second integral by parts gives

$$\int dx\;\;x_j\;\nabla_jv_j^2\;EA(t,\boldsymbol{x}) = -\int dx\,v_j^2\;EA(t,\boldsymbol{x}) = -\int dx\,(v^2(t,\boldsymbol{x}))^2A(t,\boldsymbol{x})$$

Let denote:

$$v^4A(t) = \int dx\,(v^2(t,\boldsymbol{x}))^2A(t,\boldsymbol{x}) = \;\int dx\,v^4A(t,\boldsymbol{x}) \tag{C.10.3}$$

$$u^4B(t) = \int dx\,(u^2(t,\boldsymbol{x}))^2B(t,\boldsymbol{x}) = \int dx\,u^4B(t,\boldsymbol{x}) \tag{C.10.4}$$

New variables $v^4A(t,\boldsymbol{x})$ and $u^4B(t,\boldsymbol{x})$ are proportional to forth order of velocity $v^4$ and Assets density $A(t,\boldsymbol{x})$. Equations on *XPEA(t)* and *YPEB(t)* take form:

$$\frac{d}{dt}XPEA(t) = v^4A(t) + c_{pe}\,YPEB(t)\;\;;\;\;\frac{d}{dt}YPEB(t) = u^4B(t) + c_{pe}\,XPEA(t) \tag{C.11}$$

Let introduce equations on variables $v^4A(t,\boldsymbol{x})$ and $u^4B(t,\boldsymbol{x})$ similar to equations (A.8) on Assets energy and assume that $v^4A(t)$ and $u^4B(t)$ should grow up as exponent:

$$\frac{\partial v^4A}{\partial t} + \nabla\cdot(\boldsymbol{v}v^4A) = c_u\,u^4B(t,x)\,;\;\frac{\partial u^4B}{\partial t} + \nabla\cdot(\boldsymbol{u}\,u^4B) = d_v\,v^4A(t,x) \tag{C.12.1}$$

$$\frac{d}{dt}v^4A(t) = c_{ue}\,u^4B(t)\;\;;\;\;\frac{d}{dt}u^4B(t) = c_{ve}\,v^4A(t) \tag{C.12.2}$$

$$\gamma^2{}_{vu} = c_ud_v > 0 \tag{C.12.3}$$

Equations (C.12.1-3) describe exponential growth of $v^4A(t)$ and $u^4B(t)$ alike to equations (A.10). To close system of equations let define $XVA$ and $YUB$ (C.7.3; C.7.4). For simplicity let assume that these factors are described by equations alike to (A.8; A.10):

$$\frac{\partial}{\partial t}XVA(t,\boldsymbol{x}) + \nabla\cdot(\boldsymbol{v}(t,\boldsymbol{x})XVA) = c_{xv}\,YUB(t,\boldsymbol{x}) \qquad \text{(C.13.1)}$$

$$\frac{\partial}{\partial t}YUB(t,\boldsymbol{x}) + \nabla\cdot(\boldsymbol{u}(t,\boldsymbol{x})YUB) = d_{xv}\,XVA(t,\boldsymbol{x}) \qquad \text{(C.13.2)}$$

Then equations on $XVA(t)$ and *YUB(t)* take form:

$$\frac{d}{dt}XVA(t) = c_{xv}\,YUB(t) \quad ; \quad \frac{d}{dt}YUB(t) = d_{xv}\,XVA(t) \qquad \text{(C.14)}$$

$$\gamma_{xv}{}^2 = c_{xv}d_{xv} > 0 \qquad \text{(C.15)}$$

Equations (A.5; A.10; C.5; C.8.1; C.8.2; C.9; C.11; C.12.2; C.14) describe mean squares $\overline{X^2}(t)A(t)$ and $\overline{Y^2}(t)B(t)$ and all variables that determine these factors: *XPA(t), YPB(t), EA(t), EB(t),* $XPAX^2(t)$*,* $YPBY^2(t)$*,* $X^2EA(t)$*,* $Y^2EB(t)$*, XPEA(t), YPEB(t),* $v^2EA(t), u^2EB(t)$ *XVA(t)* and *YUB(t).*We omit long but simple calculations that allow derive solution of system and present general result as

$$\overline{X^2}(t)A(t) = \overline{X^2}A(0) + \overline{X^2}A(1)\sin\omega t + \overline{X^2}A(2)\cos\omega t + \overline{X^2}A(3)\sin\omega_{pe}t + \overline{X^2}A(4)\exp\gamma_e t +$$
$$\overline{X^2}A(5)\ \exp-\gamma_e t + \overline{X^2}A(6)\ \exp\gamma_{vu}t + \overline{X^2}A(7)\ \exp-\gamma_{vu}t + \overline{X^2}A(8)\ \exp\gamma_{xv}t +$$
$$\overline{X^2}A(9)\ \exp-\gamma_{xv}t \qquad \text{(C.16)}$$

with constants $\overline{X^2}A(j)$, *j=0,..9* defined by initial value of variables (A.12; A.13; C.6; C.7.2; C.7.3; C.7.4; C.10.1; C.10.3; C.10.4) for *t=0. A(t)* has form (A.14) with exponential increment $\gamma_e$. As $\overline{X^2}(t)$ can't grow up as exponent hence it is required that

(A.12; A.13; C.6; C.7.2; C.7.3; C.7.4; C.10.1; C.10.3; C.10.4)$\gamma_e > \gamma_{vu}$ ; $\gamma_e > \gamma_{xv}$

Revenue-on-Assets mean square risk $\overline{\mathrm{Y}^2}(\mathrm{t})\mathrm{B}(\mathrm{t})$ is defined by relations similar to (C.16) with constants defined by initial value of variables (A.12; A.13; C.6; C.7.2; C.7.3; C.7.4; C.10.1; C.10.3; C.10.4).

## References

1. Schumpeter, J.A. Business Cycles. A Theoretical, Historical and Statistical Analysis of the Capitalist Process. NY McGraw-Hill Book Company, 461 pp., 1939.
2. Lucas, R.E. Methods and Problems in Business Cycle Theory, Jour.of Money, Credit and Banking, 12, (4) Part 2: Rational Expectations, 696-715, 1980
3. Zarnowitz, V. Business Cycles: Theory, History, Indicators, and Forecasting. NBER, University of Chicago Press, 1992.
4. Lucas, R.E. Understanding Business Cycles, in Estrin, S and A. Marin (ed.), Essential Readings in Economics, pp. 306-327. Macmillan, 1995.
5. Rebelo, S. Real Business Cycle Models: Past, Present, And Future, NBER, WP 11401, Cambridge, MA., 2005.
6. Kiyotaki, N. A Perspective on Modern Business Cycle Theory. Economic Quarterly, 97, (3), 195–208, 2011.
7. Jorda, O., Schularick, M., Taylor, A.M. Macrofinancial History and the New Business Cycle Facts. FRB San Francisco, WP 2016-23, 2016.
8. Huggett, M. Business Cycles: Facts and Theory. Lecture 8. Economics Department, Georgetown University, Washington DC., 2017. http://faculty.georgetown.edu/mh5/class/econ102/
9. Starr, R.M. General Equilibrium Theory, An Introduction, Cam. Univ. NY., 2011.
10. Simon, H. A. Theories of decision-making and behavioral science. The American Economic Review, 49, 253–283, 1959.
11. Cramer, C.F., Loewenstein, G., Rabin, M. (Ed) Advances in Behavioral Economics, Princeton Univ. Press, 2004.
12. Olkhov, V. On Economic space Notion, International Review of Financial Analysis, 47, 372-381, 2016. DOI-10.1016/j.irfa.2016.01.001
13. Olkhov, V. Finance, Risk and Economic space, ACRN Oxford J. of Finance and Risk Perspectives, Special Issue of Finance Risk and Accounting Perspectives, 5: 209-221, 2016.
14. Olkhov, V. Quantitative Wave Model of Macro-Finance. International Review of Financial Analysis, 50, 143-150, 2017.
15. Olkhov, V. Econophysics Macroeconomic Model, http://arxiv.org/abs/1701.06625, q-fin.EC, 2017.
16. Olkhov, V. Econophysics of Macroeconomics: "Action-at-a-Distance" and Waves, http://arxiv.org/abs/1702.02763, 2017.
17. Olkhov, V., Econophysics of Macro-Finance: Local Multi-fluid Models and Surface-like Waves of Financial Variables, arXiv:1706.01748, q-fin.EC , 2017.

18. Olkhov, V. Non-Local Macroeconomic Transactions and Credits-Loans Surface-Like Waves, http://arxiv.org/abs/1706.07758, q-fin.EC , 2017.

19. Tesfatsion, L., Judd, K. (Eds.). Handbook of computational economics. Agent-Based Computational Economics, Vol. 2. North-Holland: Elsevier, 2005.

20. Fujita, M. The Evolution Of Spatial Economics: From Thünen To The New Economic Geography, The Japanese Economic Review 61: 1-32, 2010.

21. Chane-Kon, L., Neugebauer, M., Pak, A., Trebach, K., Tsang, E. Structured Finance. Global Rating Criteria for Structured Finance CDOs, Fitch Ratings, 1-27, 2010.

22. Kraemer, N., Vazza, D. 2011 Annual U.S. Corporate Default Study And Rating Transitions, S&P: 1-96, 2012.

23. Metz, A., Cantor, R. Introducing Moody's Credit Transition Model, Moody's Investor Service, 1-26, 2007.

24. Landau, L.D., Lifshitz, E.M. Fluid Mechanics, Pergamon Press Ltd., NY., 1987.